\documentclass[final,3p,times]{elsarticle}

\usepackage{graphics}
\usepackage{amsmath}
\usepackage{latexsym}

\usepackage[T1]{fontenc}
\usepackage{amssymb}
\usepackage{natbib}
\usepackage{latexsym}
\usepackage{color,graphicx}  
\usepackage[bottom]{footmisc}
\usepackage{fancyvrb}  
\usepackage{times}
\usepackage{hyperref}
\usepackage{amsthm}
\usepackage{cancel}
\usepackage{ulem}

\newcommand{\mand}   {\quad \mathrm{and} \quad}

\newcommand\eqnref[1]{(\ref{#1})}
\newcommand\figref[1]{Fig.~\ref{#1}}

\newcommand\sectref[1]{Section~\ref{#1}}

\newcommand{\Sobolev}   {{H}}
\newcommand{\Hilbert}   {{H}}
\newcommand{\Lebesgue}   {{L}}

\newcommand{\Omegarm}   {\mathrm{\Omega}}

\newcommand{\vphirm}   {\mathrm{\varphi}}

\newcommand{\Einc}    {E_{\mathrm{inc}}}
\newcommand{\Hinc}    {H_{\mathrm{inc}}}
\newcommand{\Eref}    {E_{\mathrm{ref}}}
\newcommand{\Href}    {H_{\mathrm{ref}}}
\newcommand{\Etrs}    {E_{\mathrm{trs}}}
\newcommand{\Htrs}    {H_{\mathrm{trs}}}
\newcommand{\winc}   {\vphirm_{\mathrm{inc}}}
\newcommand{\wref}   {\vphirm_{\mathrm{ref}}}
\newcommand{\wtrs}   {\vphirm_{\mathrm{trs}}}
\newcommand{\wgen}   {\vphirm}

\makeatletter
\def\Eqlfill@{\arrowfill@\Relbar\Relbar\Relbar}
\newcommand{\extendEql}[1][]{\ext@arrow 0359\Eqlfill@{#1}}
\makeatother

\usepackage[usenames,dvipsnames]{xcolor}

\journal{Journal of Computational Physics}

\begin{document}
\begin{frontmatter}
%
%
\title{Discontinuous Galerkin Methods with Trefftz Approximation}
\author[gsc,temf]{Fritz Kretzschmar\corref{cor}}\ead{kretzschmar@gsc.tu-darmstadt.de}
\author[ifc]{Sascha M. Schnepp}\ead{schnepps@ethz.ch}
\author[dece]{Igor Tsukerman}\ead{igor@uakron.edu}
\author[temf]{Thomas Weiland}\ead{thomas.weiland@temf.tu-darmstadt.de}
\address[gsc]{Graduate School of Computational Engineering,  Technische Universitaet Darmstadt, Dolivostrasse 15  , 64293 Darmstadt, Germany}
\address[temf]{Institut fuer Theorie Elektromagnetischer Felder, Technische Universitaet Darmstadt, Schlossgartenstrasse 8  , 64289 Darmstadt, Germany}
\address[ifc]{Laboratory for Electromagnetic Fields and Microwave Electronics, ETH Zurich, Gloriastrasse 35,  8092 Zurich, Switzerland}
\address[dece]{Department of Electrical \& Computer Engineering, The University of Akron, Akron, Ohio 44325-3904,USA}
\cortext[cor]{Corresponding Author}
%
%
%
\begin{abstract}
We present a novel Discontinuous Galerkin Finite Element Method for wave propagation problems. The method employs space-time Trefftz-type basis functions that satisfy the underlying partial differential equations and the respective interface boundary conditions exactly in an element-wise fashion. 
The basis functions can be of arbitrary high order, and we demonstrate spectral convergence in the $\Lebesgue_2$-norm. In this context, spectral convergence is obtained with respect to the approximation error in the entire space-time domain of interest, i.e. in space and time simultaneously. 
Formulating the approximation in terms of a space-time Trefftz basis makes high order time integration an inherent property of the method and clearly sets it apart from methods, that employ a high order approximation in space only. 
%
%
\end{abstract}
%
%
\end{frontmatter}
%
%
\section{Introduction}\label{sec:intro} 
%
In the context of Finite Element Methods (FEM) there exist two main ways of improving the numerical accuracy. First, a computational mesh on
which a problem is approximated can be refined ($h$-refinement). Secondly, the order $p$ of the approximating polynomials can be increased ($p$-refinement). 
Both strategies, their combinations ($hp$-refinement) and their connection to the level of regularity of the solutions have been extensively studied (see e.g.~\cite{Babuska92,Schwab98}).

Even higher accuracy can be obtained by applying problem-specific high order basis functions. 
We pursue this approach in the framework of the Discontinuous Galerkin Finite Element
Methods (DG-FEM) \cite{Reed1973,LeSaintRaviart1974,Cockburn2001,Fezoui2005}, as it allows
an almost unrestricted choice of bases.
To date, there exist only few works on DG and DG-type methods using a problem
specific basis in the frequency-domain, in particular the Ultra Weak Variational Formulation (UWVF)~\cite{Cessenat98,Monk02,Huttunen:2007tq} and the
Plane Wave Discontinuous Galerkin Method (PWDG)~\cite{Hiptmair:2011p1917,Moiola:2011io} which has been shown to be a generalization of the former one. In the time-domain, we are aware of~\cite{Petersen:2009id} only. However, this approach differs significantly from our work as will be detailed below.
Most commonly however, generic polynomials are employed, which does not lead to the
optimal approximation accuracy for a given polynomial order and mesh size. 

In this article we present a highly accurate type of DG-FEM for time-domain applications.
A distinguishing new attribute of the method is the use of Trefftz-type basis
functions~\cite{trefftz1926,Runge89,Jirousek97}, which, by definition, exactly satisfy the underlying
partial differential equations and the respective interface boundary conditions in an element-wise fashion. The method is, hence, 
a Discontinuous Galerkin Trefftz Finite Element Method (DGT-FEM). 

We consider space- and time-dependent equations. As Trefftz functions are required to solve the equations exactly, at least in a local sense, they necessarily have to depend on time as well. Consequently, the DGT-FEM has to be formulated as a space-time FEM~\cite{Hulbert:1990uu,Vandervegt:2002ba}.
In a space-time method there is no formal distinction between the spatial dimensions and
the temporal one. Finite element spaces are given as $(d+1)$-dimensional functional spaces,
where $d$ denotes the number of spatial dimensions.

Working in a space-time setting has two immediate implications: 
first, high order time integration is obtained in a straightforward and consistent manner
by increasing the order of the approximating polynomials, and second, convergence
rates describe the approximation error not only in space but in space-time. It is shown below
that spectral convergence of the global approximation error in the space-time domain of interest
is obtained under $p$-enrichment for sufficiently smooth data.
This is in contrast to DG methods based on a discretization of space, where spectral convergence is 
attained with respect to the spatial discretization error only. We also show that the presence of inhomogeneous 
elements (i.e elements filled with different materials) does not affect the convergence rate if an appropriate 
local Trefftz basis is used.

The remainder of this article is organized as follows. 
In \sectref{sec:Model-1D} we derive the weak space-time DG formulation
of time-dependent electromagnetic waves. The derivation is restricted
to the $(1+1)$-dimensional case. A generalization to higher spatial dimensions
is outlined below and will be implemented in the near future. \sectref{sec:Trefftz-approximations}  introduces the space-time Trefftz basis
functions that by necessity have to be vector-valued, with the electric and magnetic field as components.
Trefftz bases are derived for homogeneously filled grid cells as well as for partially filled cells containing two different materials. 
Results obtained from numerical experiments are presented in \sectref{sec:results}.
First, we demonstrate the physical correctness of the results. Then we investigate the convergence rates of the method both in space and time.
Finally, we show that the method outperforms well-established methods such as the Finite Difference Time Domain Method (FDTD)
and the DG method with leapfrog time stepping (DGL) in terms of efficiency. \sectref{sec:outlook} concludes the paper and provides an outlook
to the ongoing work of extending the method to higher dimensions in space.
%
%
\section{Time-Dependent One-Dimensional Waves}\label{sec:Model-1D}
%
This section contains a derivation of a weak space-time DG formulation of Maxwell's equations using vectorial basis functions. 
The resulting equations describe the propagation of electromagnetic waves in one spatial dimension. The procedure developed 
for electromagnetic waves in this section is applicable to waves of any nature. However, we shall focus on electromagnetic analysis for definiteness.
%
\subsection{Differential Formulation of the Problem in Continuum} \label{sec:Differential-formulation}
%
For a wave propagating in a given direction $x$ with one-component fields $E\equiv E_y$ and $H\equiv H_z$,
the system of Maxwell's equations simplifies to
\begin{align} \label{eqn:1D-Maxwell}
    \partial_x E + \partial_t \big(\mu H \big) = 0 \quad \mathrm{and} \quad \partial_x H + \partial_t \big(\epsilon E \big) = J,
\end{align}
where only the time-dependent equations are considered. Here, $\epsilon$ and $\mu$ are given material parameters that can in general be functions of both position and time (in electrodynamics, these are the dielectric
permittivity and the magnetic permeability, respectively). $J$ is a given source that can also depend on space $x$ and time $t$. The electric field $E \in \Sobolev^1(\Omegarm)$ and the 
magnetic field $H \in \Sobolev^1(\Omegarm)$ are defined in the Sobolev space
\begin{align*}
 \Sobolev^1 (\Omegarm) \equiv \big{\{} u \in L_2 (\Omegarm): \nabla u \in L_2 (\Omegarm) \big{\}},
\end{align*}
with $L_2 (\Omegarm)$ being the space of square integrable functions over the space-time domain of interest
\begin{align*}
 \Omegarm \equiv \mathbf{I}_{t} \times \mathbf{I}_{x} \subset \mathbb{R}^2 \qquad \mathrm{with} \qquad
 \mathbf{I}_{t} \equiv [t_{\mathrm{min}}; t_{\mathrm{max}}] \subset \mathbb{R} \quad 
\mathrm{and} \quad \mathbf{I}_{x} \equiv [x_{\mathrm{min}}; x_{\mathrm{max}}] \subset \mathbb{R}.
\end{align*}
For highlighting the generality of the following derivation, we rewrite equations \eqnref{eqn:1D-Maxwell} in a coordinate-independent divergence form
\begin{align} \label{eqn:1D-Maxwell-div-form} 
\left(\begin{array}{c}
  \partial_t \\ \partial_x
 \end{array}\right)^{\mathrm{T}} \cdot
\left(\begin{array}{cc}
  0 &\mu \\1&0
 \end{array}\right) \cdot
\left(\begin{array}{c}
  E \\ H
 \end{array}\right)& \,=\, 0,\nonumber 
\\ \\
 \left(\begin{array}{c}
  \partial_t \\ \partial_x
 \end{array}\right)^{\mathrm{T}} \cdot
\left(\begin{array}{cc}
  \epsilon &0 \\0&1
 \end{array}\right) \cdot
\left(\begin{array}{c}
  E \\ H
 \end{array}\right)& \,=\, J. \nonumber
\end{align}
By defining the material and differential operators
\begin{align*}
 \mathrm{\eta}_\epsilon \equiv \left(\begin{array}{cc}
  \epsilon &0 \\0&1
 \end{array}\right), \, \,
\mathrm{\eta}_\mu \equiv
\left(\begin{array}{cc}
  0 &\mu \\ 1&0
 \end{array}\right) \, \, \mathrm{and} \, \, \mathbf{\nabla} \equiv\left(\begin{array}{c}
  \partial_t \\ \partial_x
 \end{array}\right), \nonumber
\end{align*}
we can further contract equations \eqnref{eqn:1D-Maxwell-div-form} into the more compact form
\begin{align} \label{eqn:1D-Maxwell-compact-form}
  \mathbf{\nabla}^{\mathrm{T}} \cdot \mathrm{\eta}_\mu \cdot \mathbf{F} = 0 \mand \mathbf{\nabla}^{\mathrm{T}} \cdot \mathrm{\eta}_\epsilon \cdot  \mathbf{F} = J.
\end{align}
Here $\mathbf{F} \equiv (E,H)^{\mathrm{T}} \in \Sobolev^1(\Omegarm) \times \Sobolev^1(\Omegarm)$ is the field vector defined in the vectorial Sobolev space. Equations 
\eqnref{eqn:1D-Maxwell-compact-form} are subject to initial as well as boundary conditions.
%
\subsection{The Weak Form}\label{sec:Weak-form}
%
In order to obtain a weak formulation of \eqref{eqn:1D-Maxwell-compact-form}, we
first cast it into the more abstract form
\begin{align}\label{eqn:1D-Maxwell-abstract-form}
\mathcal{L} \, \mathbf{F} = \mathbf{J},
\end{align}
where we used the source vector $\mathbf{J} \equiv (0,J)^{\mathrm{T}} \in \Lebesgue_2(\Omegarm) \,\times\, \Lebesgue_2(\Omegarm)$ 
and the differential Maxwell operator $\mathcal{L}: \Sobolev^1(\Omegarm) \times \Sobolev^1(\Omegarm) \rightarrow \Lebesgue_2(\Omegarm) \times \Lebesgue_2(\Omegarm)$ 
defined as $\mathcal{L} \equiv (\mathbf{\nabla}^{\mathrm{T}} \mathrm{\eta}_\epsilon,\mathbf{\nabla}^{\mathrm{T}} \mathrm{\eta}_\mu)$. We then produce a weak formulation, 
by composing the inner product of both sides of \eqnref{eqn:1D-Maxwell-abstract-form} with a test function $\mathbf{v} \equiv (v^E,v^H)^{\mathrm{T}} \in \Sobolev^1(\Omegarm) \times \Sobolev^1(\Omegarm)$ as
\begin{align} \label{eqn:weak-form-abstract}
   \big[ \mathcal{L} \, \mathbf{F} ,\, \mathbf{v} \big]_{\Omegarm} \,=\, \big[\mathbf{J},\mathbf{v} \big]_{\Omegarm},
\end{align}
where $[ \cdot , \cdot ]_{\Omegarm}$ is the inner product in $\Lebesgue_2(\Omegarm) \times \Lebesgue_2(\Omegarm)$ defined as
\begin{align*}
  \big[\mathbf{F} , \mathbf{v} \big]_{\Omegarm} \equiv \big(E ,v^E \big)_{\Omegarm} + \big(H ,v^H \big)_{\Omegarm},
\end{align*}
and $( \cdot , \cdot )_{\Omegarm}$ is the standard $\Lebesgue_2$ inner product in $\Omegarm$.
Casting the weak formulation \eqnref{eqn:weak-form-abstract} back into a more explicit form yields
\begin{align} \label{eqn:weak-form-explicit}
  \big(\mathbf{\nabla}^{\mathrm{T}} \mathrm{\eta}_\epsilon \mathbf{F} ,v^E \big)_{\Omegarm} + \big(\mathbf{\nabla}^{\mathrm{T}} \mathrm{\eta}_\mu \mathbf{F} ,v^H \big)_{\Omegarm} = \big(J ,v^H \big)_{\Omegarm}.
\end{align}
%
%
\subsection{The DG Formulation}\label{sec:DG-formulation}
%
To obtain a DG formulation of equation \eqnref{eqn:weak-form-explicit}  the expression is transformed using integration by parts. This yields
\begin{align}\label{eqn:weak-form-int-by-parts}
  \int_{\partial \Omegarm} v^{E} \big( \mathbf{\eta}_\epsilon  \mathbf{F} \big) \cdot \mathbf{n} \,\mathrm{d}\Omegarm
  + \int_{\partial \Omegarm} v^{H} \big( \mathbf{\eta}_\mu  \mathbf{F} \big) \cdot \mathbf{n} \,\mathrm{d}\Omegarm
  - \big( \mathbf{\nabla}^{\mathrm{T}}  \cdot v^{E},\mathrm{\eta}_\epsilon \mathbf{F}\big)_{\Omegarm}
  -\big(  \mathbf{\nabla}^{\mathrm{T}} \cdot v^{H} , \mathrm{\eta}_\mu \mathbf{F}  \big)_{\Omegarm} 
  = \big(J ,v^H\big)_{\Omegarm},
\end{align}
where $\mathbf{n} = (n_t,n_x)^{\mathrm{T}}$ is the unit outward normal on the space-time domain
boundary $\partial \Omegarm$. More specifically, we consider a partition $\Omegarm_{h}$  of the space-time domain 
of interest $\Omegarm$ on a Cartesian grid (not necessarily uniform) into $K=N \cdot M$ cells. 
Here $N$ denotes the total number of cells in the temporal direction whereas $M$ denotes the total 
number of cells in the spatial direction. Each cell with index $k=(n,m)$ is denoted by $\Omegarm_k$ where $n$ 
is the temporal index and $m$ the spatial index. The components of the test functions $v^{E}$ and $v^{H}$ are 
defined cell wise in their discretized Sobolev spaces
\begin{align*}
\Sobolev^1 (\Omegarm_h) \equiv \bigg\{ v^i \, \in \,  \Lebesgue_2 (\Omegarm): \, \forall \Omegarm_k\, \in\, \Omegarm_h,\, v^i_{{|} \Omegarm_k} \,\in\, \Hilbert^1 (\Omegarm_k) \,\bigg\} \quad \mathrm{with} \quad i=E,H. 
\end{align*}
The vectorial test function is therefore defined in $\mathbf{v} \in \Sobolev^1 (\Omegarm_h) \times \Sobolev^1 (\Omegarm_h)$.

\begin{figure}[!ht]
\centering
\includegraphics[width=1.0\textwidth]{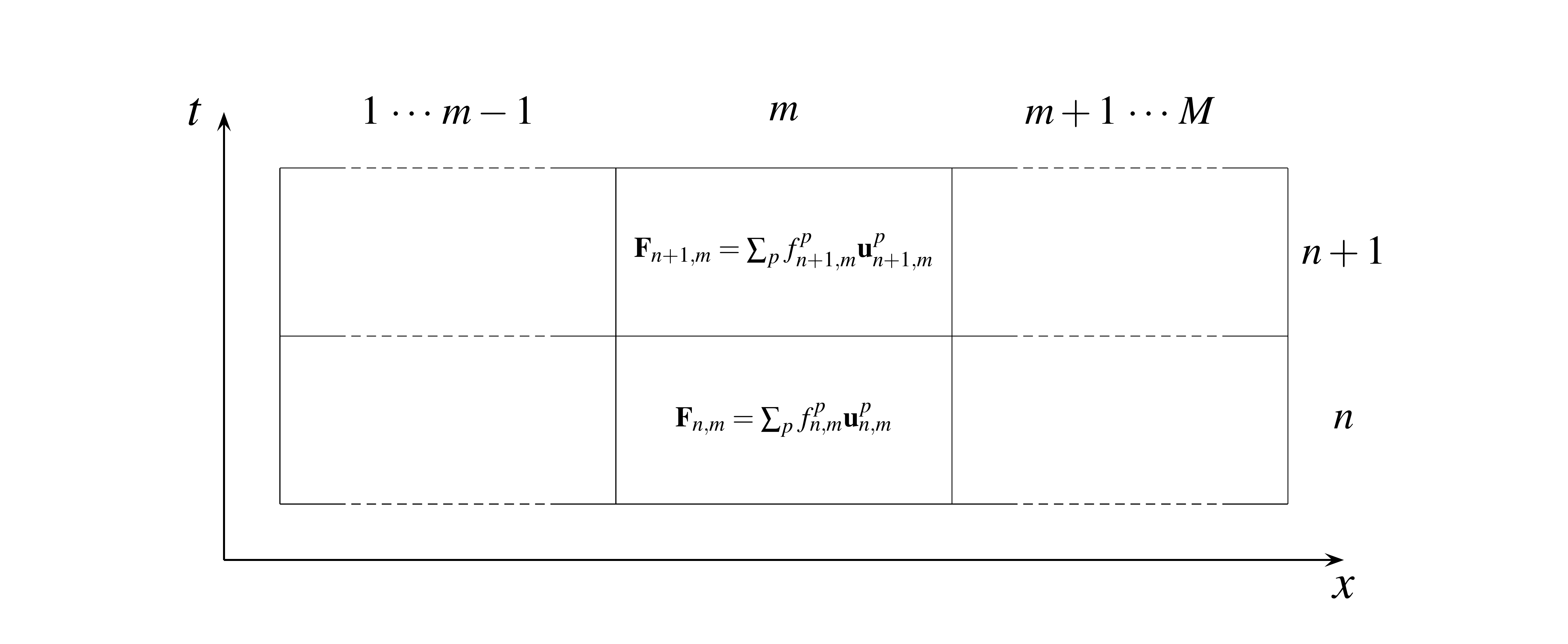}
 \caption[fig:two-time-layers]{Space-time cells for two subsequent time layers. 
 \label{fig:two-time-layers}}
\end{figure}

To generate a time-stepping scheme on $\Omegarm_h$, we will consider
only two spatial layers of grid cells, defined at two consecutive time slabs labeled
$n$ and $n + 1$ (see. \figref{fig:two-time-layers}). It is inherent to our method that 
space and time inside one time slab are treated on an equal footing in the discretization procedure. 
We will apply \eqnref{eqn:weak-form-int-by-parts} to a single grid cell $\Omegarm_{m,n+1}$ at time-level $n + 1$ rather than $n$, as our focus is on implicit schemes.
The field vector $\mathbf{F}$ is separately approximated within each cell using a linear combination of basis functions $\mathbf{u}$, viz.
\begin{align}\label{eqn:EH-eq-cm-psim}
    \mathbf{F}_{n,m} (x, t)  = 
    \sum\nolimits_p  f^p_{n,m}  \mathbf{u}^p_{n,m} (x, t),
\end{align}
where each basis function is vectorial, with the two components
$\mathbf{u}^p_{n,m} \equiv (u^{E,p}_{n,m}, u^{H,p}_{n,m} )^{\mathrm{T}} \in \Sobolev^1(\Omegarm_{n,m})\, \times \,\Sobolev^1(\Omegarm_{n,m})$.
Note that the approximation is of order $p$ in space and time while applying the same
number of Degrees of Freedom (DoF) as a corresponding approximation of space only.

In contrast to most approaches, we are interested in employing Trefftz bases rather than the standard piecewise-polynomial approximations. However, in this part of the derivation we keep a general form of basis functions and implement the Trefftz basis in \sectref{sec:Trefftz-approximations}.

Once a set of basis functions over the cell is chosen, the integral over $\Omegarm_{n+1,m}$ directly leads to the entries of the stiffness matrices  
\begin{align*}
 \mathbf{S}^{p,q\,\mathrm{(space)}}_{n+1,m} =\int \int  \bigg( u^{E,p} \, \big( \partial_x u^{H,q} \big) + u^{H,p} \, \big( \partial_x u^{E,q} \big) \bigg)_{n+1,m}  \,dx \, dt,
\end{align*}
\begin{align*}
 \mathbf{S}^{p,q\,\mathrm{(time)}}_{n+1,m} =   \int \int  \bigg( \epsilon u^{E,p} \, \big( \partial_t u^{E,q} \big) + \mu u^{H,p} \, \big( \partial_t u^{H,q} \big) \bigg)_{n+1,m}  \,dx \, dt.
\end{align*}
On the element level, the treatment of the integrals over the element boundary $\Gamma_{n+1,m}$ is more nuanced than in the global problem \eqref{eqn:weak-form-int-by-parts}.
These integrals represent surface fluxes that in the continuous problem must be the same for any two adjacent grid cells.
In DG methods basis and test functions are defined with a strictly element-wise compact support. Therefore,
the approximation is continuous within each element but discontinuous across any cell interfaces. At the interfaces
the global approximation is double-valued at every point and interface fluxes from adjacent grid cells can differ.
Flux continuity can be numerically imposed via a penalty term or, as is commonly done
in DG methods, by combining weighted numerical approximations of each flux in the two adjacent cells.
More specifically, we are using flux expressions that are centered in space and upwind in time. 
Let us write out all the fluxes in terms of basis and test functions of maximum order $P_m$ and $Q_m$, respectively, 
in a given space-time cell $\Omegarm_{n,m}$.  We denote evaluations of the basis functions on the left and right hand 
cell boundary using left and right arrows ($\leftarrow , \rightarrow$), respectively.
In temporal direction we indicate evaluations of the basis functions at the upper (i.e.~towards the next time level) and lower
boundary using up and down pointing arrows ($\uparrow, \downarrow$). The element boundary $\Gamma_{n+1,m}$ is dissected into the 
four element edges. The space flux in cell $\Omegarm_{n+1,m}$ is composed from three terms. Cell $\Omegarm_{n+1,m}$ itself contributes 
\begin{align*}
\mathbf{L}^{q,p\, \mathrm{(space)}}_{n+1,m\,, \bullet} = \frac{1}{2} \int  \bigg( u_{m}^{E,p,\rightarrow} u_{m}^{H,q,\rightarrow} + u_{m}^{H,p,\rightarrow} u_{m}^{E,q,\rightarrow} \bigg)_{n+1}  \,dt
 -  \frac{1}{2} \int \bigg( u_{m}^{E,p,\leftarrow} u_{m}^{H,q,\leftarrow} + u_{m}^{H,p,\leftarrow} u_{m}^{E,q,\leftarrow} \bigg)_{n+1}  \,dt.
\end{align*}
In addition, the neighboring cell on the right $\Omegarm_{n+1,m+1}$ contributes with
\begin{align*}
\mathbf{L}^{q,p\,\mathrm{(space)}}_{n+1,m\,, \rightarrow} = \frac{1}{2} \int  \bigg( u_{m+1}^{E,p,\leftarrow} u_{m}^{H,q,\rightarrow} + u_{m+1}^{H,p,\leftarrow} u_{m}^{E,q,\rightarrow} \bigg)_{n+1}  \,dt,
\end{align*}
and to the neighboring cell on the left $\Omegarm_{n+1,m-1}$ with
\begin{align*}
 \mathbf{L}^{q,p\,\mathrm{(space)}}_{n+1,m\,, \leftarrow  } =  - \frac{1}{2} \int  \bigg( u_{m-1}^{E,p,\rightarrow} u_{m}^{H,q,\leftarrow} + u_{m-1}^{H,p,\rightarrow} u_{m}^{E,q,\leftarrow} \bigg)_{n+1}  \,dt.
\end{align*}
The time flux of cell $\Omegarm_{n+1,m}$ has only two contributions from cells $\Omegarm_{n+1,m}$ and $\Omegarm_{n,m}$ 
\begin{align*}
\mathbf{L}^{q,p\, \mathrm{(space)}}_{n+1,m\,, \bullet} &= \quad \int  \bigg( \epsilon u_{n+1}^{E,p,\uparrow} \, u_{n+1}^{E,q,\uparrow}  + \mu u_{n+1}^{H,p,\uparrow}  \, u_{n+1}^{H,q,\uparrow} \bigg)_m  \,dx, \\ \mathbf{L}^{q,p\,\mathrm{(time)}}_{n+1,m\, , \downarrow} &=
-\int  \bigg(  \epsilon u_{n}^{E,p,\uparrow} \, u_{n+1}^{E,q,\downarrow}  +  \mu u_{n}^{H,p,\uparrow} \, u_{n+1}^{H,q,\downarrow}  \bigg)_m  \,dx.
\end{align*}
We can now rewrite Maxwell's equations in a fully discretized matrix form, which reads
\begin{align*}
\bigg( \mathbf{L}^{q,p\,\mathrm{(time)}}_{n+1,m\,, \bullet} + \mathbf{L}^{q,p\,\mathrm{(space)}}_{n+1,m\,, \bullet}  - \mathbf{S}^{q,p\,\mathrm{(time)}}_{n+1,m} - \mathbf{S}^{q,p\,\mathrm{(space)}}_{n+1,m}  \bigg) \, f^{p}_{n+1,m}  + \mathbf{L}^{q,p\,\mathrm{(space)}}_{n+1,m\,, \rightarrow}f^{p}_{n+1,m+1} + \mathbf{L}^{q,p\,\mathrm{(space)}}_{n+1,m\,, \leftarrow} f^{p}_{n+1,m-1} = \mathbf{L}^{q,p\,\mathrm{(time)}}_{n+1,m\,, \downarrow} f^{p}_{n,m}.
\end{align*}
We further combine the stiffness and flux matrix expressions on the left and right hand sides into two global matrices 
$\mathbf{A}$ and $\mathbf{B}$ and formally obtain the implicit time update equation
\begin{align}\label{eqn:matrix-maxwell}
\mathbf{A} f_{n+1} = \mathbf{B} f_{n}
\end{align}
for progressing from time layer $n$ to a subsequent time layer $n+1$. 
%
\section{Approximation with Trefftz Functions}\label{sec:Trefftz-approximations}
%
Trefftz basis functions are a set of problem-specific functions, which satisfy the underlying
partial differential equations exactly within a grid cell. As they include features
of the exact solution, Trefftz methods often yield very accurate numerical solutions.
In this section we construct a Trefftz basis for electromagnetic wave propagation problems as stated in \sectref{sec:Model-1D}. 
For some scenarios, such as the propagation of a plane wave in a vacuum, Trefftz basis functions are relatively easy to 
construct~\cite{Tsukerman06, Tsukerman-book07}. However, we also address the case of partially filled elements, where the derivation of locally exact functions is more involved.
%
\subsection{Trefftz Functions for (1+1)-Dimensional Maxwell's Equations}\label{sec:Trefftz-Basis}
%
In the following we apply Maxwell's equations~\eqref{eqn:1D-Maxwell} in natural units, with the permittivity and permeability of vacuum set to one. As a consequence, the speed of light in vacuum, $v_0$, is one as well. 
We omit cell indices (i.e. $k$, $n$ and $m$) for the sake of simplicity from now on.

Traveling waves of the form $\varphi (x,t) = f(x\mp vt)$, where $f$ is an arbitrary sufficiently smooth function and $v$ the local speed of light, are well known to be solutions of the wave equation.
Motivated by this, it is reasonable to include transport polynomials of the general form
\begin{align}\label{eqn:transport-basis}
  \mathbf{u}^{p,\pm} \, = \, 
  \left(\begin{array}{c}
  u^{E,p,\pm}\\ \\u^{H,p,\pm}
  \end{array}\right) \, = \,
  \left(\begin{array}{c}
  \pm \big( x \mp v \, t \big)^p \\ \\ \frac{1}{Z} \big( x \mp v \, t \big)^p
  \end{array}\right),
\end{align}  
in the basis. The first and second component, $u^E$ and $u^H$, of any basis function in \eqnref{eqn:transport-basis} 
are representing the electric and magnetic field, respectively.  The material parameters 
$Z = \sqrt{\mu  / \epsilon}$ and $v = 1 / \sqrt{\mu \epsilon}$~
(i.e. intrinsic impedance and local speed of light respectively) enter the basis functions directly. For each order $p$, 
there are two waves in the basis: one moving leftward (with opposite signs of $E$ and $H$) and the other one rightward (equal sign of $E$ and $H$).
We note that using separate scalar bases for approximating the electric and magnetic
field individually does not yield a Trefftz method. Any individual basis function or a combination
of them has to solve the governing equations.
In the two field formulation that we apply, this is not guaranteed with separate scalar basis functions as
can easily be seen by inserting transport polynomials of order $r$ and $s$ with $r \not = s$ into \eqnref{eqn:1D-Maxwell}.
The approximation of the field vector $\mathbf{F}$ is obtained as
\begin{align}\label{eqn:transport-basis-field}
  \mathbf{F} \, = \, \sum_{p=0}^{P_{\mathrm{max}}} \, \bigg( \sum_{\mathrm{sign}=+,-} \, f^{p,\mathrm{sign}} \, \, \mathbf{u}^{p,\mathrm{sign}} \, \bigg) \, \,.
\end{align}
\begin{figure}[!ht]
\centering
\includegraphics[width=0.6\textwidth]{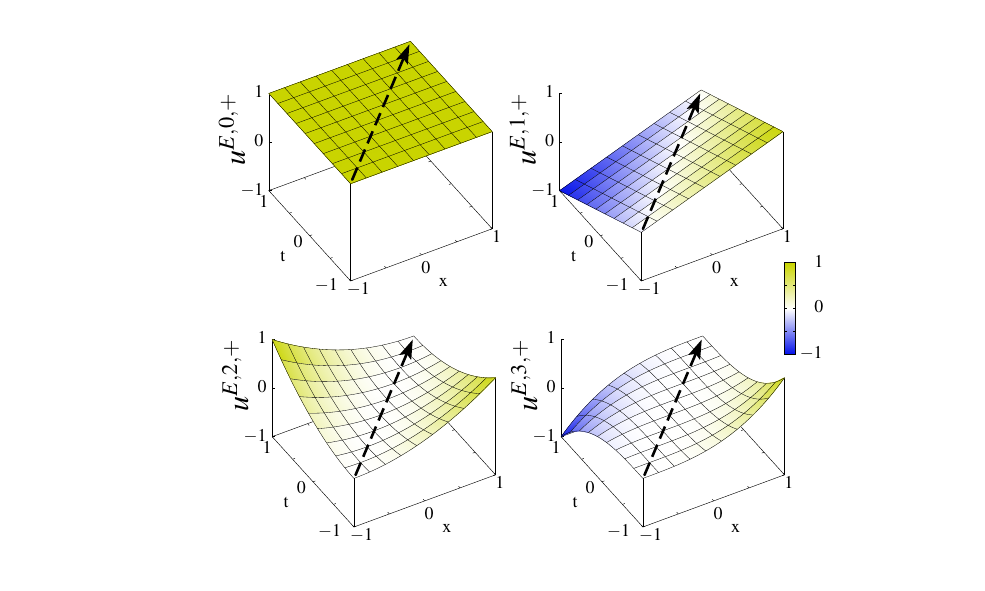}
 \caption[fig:transport-basis]{Transport polynomials of orders $p=0,1,2,3$ in a vacuum (i.e. $\epsilon=1$ and $\mu=1$) to describe waves traveling rightwards.} \label{fig:transport-basis}
\end{figure}
%
%

\begin{figure}[!ht]
\centering
\includegraphics[width=0.6\textwidth]{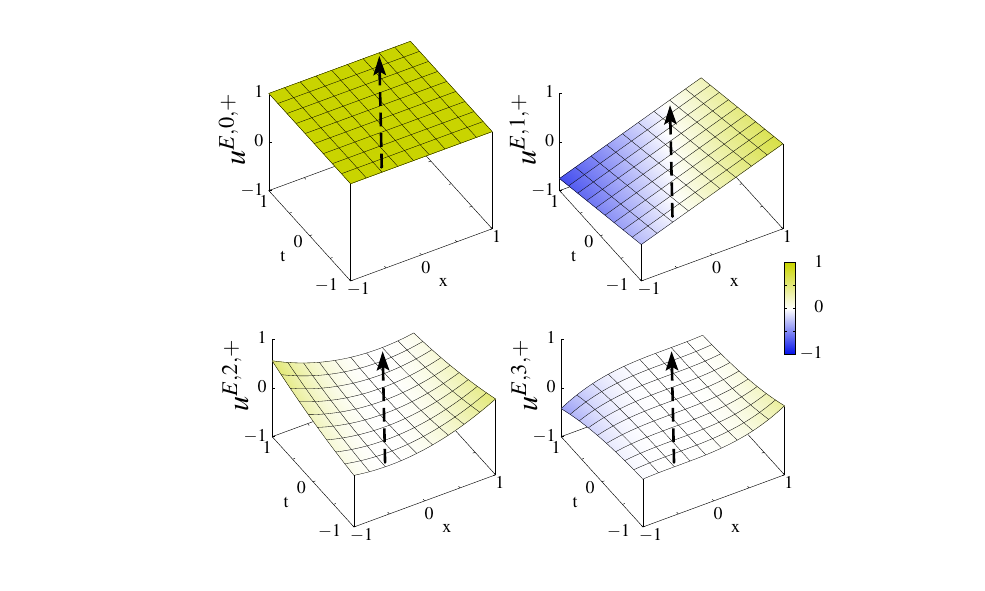}
 \caption[fig:transport-basis-medium]{Transport polynomials of orders $p=0,1,2,3$ in a medium with $\epsilon=4$ and $\mu=1$; corresponding to waves traveling rightwards.} \label{fig:transport-basis-medium}
\end{figure}

In Figs.~\ref{fig:transport-basis} and \ref{fig:transport-basis-medium} we plot the basis functions for two different media in the master element.
In both figures the cell size in spatial and temporal direction is two, however, while $\epsilon = \mu = 1$ representing  
a vacuum is used in Fig.~\ref{fig:transport-basis}, Fig.~\ref{fig:transport-basis-medium} represents the situation for a dielectric with $\epsilon = 4$ and $\mu = 1$. 
In the first case, the direction of transport is along the cell diagonal, whereas in the second case, the direction of transport is rotated away from the diagonal to an angle of 22.5 degrees.

As mentioned above, Peterson et al.~\cite{Petersen:2009id} deal with a DG time-domain method based on transport polynomials as well.
However, they discretize the second order wave equation, resulting in a weak formulation where both trial
and test functions occur in the form of derivatives only. Therefore, all constant terms vanish
rendering the stiffness matrix singular and requiring the application of regularization techniques.
A reconstruction procedure that ensures point wise inter element continuity is used to define the element average, 
which is unnatural in the context of DG methods and makes the implementation of material interfaces much harder, 
if possible at all in higher dimensions. inter-element continuity is weakly imposed by means of Lagrange multipliers 
instead of numerical fluxes. Apart from being an unusual approach within a DG method, the multipliers are additional 
degrees of freedom, which have to be solved for.
%
\subsection{Trefftz Basis Functions for Partially Filled Cells} \label{sec:Trefftz-material-interfaces}
%
In the following, a Trefftz basis is derived for partially filled cells.
To this end, we consider a space-time cell $[-\Delta_x /2, \Delta_x /2] \times [-\Delta_t /2, \Delta_t /2]$ with the origin placed at the center.
Let there be a material interface at some point $x = x_0$ inside this cell; the medium velocities and intrinsic impedances on the left and the right side of the interface being $v_{1,2}$ and $Z_{1,2}$, respectively.
We look for a Trefftz function corresponding, physically, to a traveling incident wave (`inc') that is coming from the left 
and being reflected (`ref') / transmitted (`trs') at the interface. The fields in the considered cell are
\begin{align*}
\Einc &\,=\, \winc \left(x - v_1 t\right);  \qquad &\Hinc \,=\, \frac{1}{Z_1} \winc \left(x - v_1 t\right),\\\nonumber
\Eref &\,=\, \wref \left(x + v_1 t\right);  \qquad &\Href \,=\, \frac{1}{Z_1} \wref \left(x + v_1 t\right),\\\nonumber 
\Etrs &\,=\, \wtrs \left(x - v_2 t\right);  \qquad &\Htrs \,=\, \frac{1}{Z_2} \wtrs \left(x - v_2 t\right).\nonumber 
\end{align*}
At the interface, these fields must be continuous
\begin{align*}
 \winc \left(x_0 - v_1 t\right) - \wref \left(x_0 + v_1 t\right) &= \wtrs \left(x_0 - v_2 t\right),\\\nonumber
\\\nonumber
\frac{1}{Z_1} \left(\winc \left(x_0 - v_1 t\right) + \wref \left(x_0 + v_1 t\right)\right) &= \frac{1}{Z_2} \wtrs \left(x_0 - v_2 t\right).\nonumber
\end{align*}
Solving for the reflected and transmitted waveforms, we obtain
\begin{align*}
\wref \left(x_0 + v_1 t\right) &= R \winc \left(x_0 - v_1 t\right) \quad \mathrm{with} \quad R = \frac{Z_1-Z_2}{Z_1+Z_2},\\\nonumber
\wtrs \left(x_0 - v_2 t\right) &= T \winc \left(x_0 - v_1 t\right) \quad \mathrm{with} \quad T = \frac{Z_2}{Z_1+Z_2}.\nonumber
\end{align*}
The reflection and transmission coefficients $R$ and $T$ are thus expressed in the same way as in the frequency domain,
which should be expected for frequency-independent material parameters. What remains is to find these waveforms at an 
arbitrary position $x$ rather than just at $x_0$. We therefore express the outgoing waveforms in terms of the 
incoming waveform which yields
\begin{align*}
&\wref \left(x + v_1 t\right) = \wref \left(x_0 + v_1 t + (x-x_0)\right)  \equiv \wref \left(x_0 + v_1 t^\prime\right)\\\nonumber 
& =  R \winc \left(x_0 - v_1 t^\prime \right) =  R \winc \left(2 x_0 - x - v_1 t\right),
\end{align*}
for the reflected waves and
\begin{align*}
\wtrs \left(x - v_2 t\right) = \wtrs \left(x_0 - v_2 t + (x-x_0) \right) \equiv \wtrs \left(x_0 - v_2 t^{\prime\prime}\right) \\\nonumber 
 = T \winc \left(x_0 - v_1 t^{\prime\prime} \right) =  T \winc \left(\left(1-\frac{v_1}{v_2}\right)x_0 + \frac{v_1}{v_2}x - v_1 t\right),
\end{align*}
for the transmitted waves. Here the time variables of the reflected and transmitted waveforms contain an advance or delay respectively
\begin{align*}
 t^\prime = t + \frac{(x-x_0)}{v_1} \mand t^{\prime\prime} = t - \frac{(x-x_0)}{v_2}.
\end{align*}
A set of Trefftz functions can now be obtained for any order $p$ by setting $\Einc (\wgen) = \wgen^p$ and $\Hinc (\wgen) = \wgen^p/Z$. An additional 
completely similar set can be generated for incident waves impinging on the interface boundary from the right.
%
\section{Results} \label{sec:results}
%
In this section we show numerical results obtained for different scenarios with
the DGT-FEM as introduced above. These include simulations of wave propagation in a vacuum,
and two scenarios with a material interface. In the first case the interface is located in between two elements
and in the second one inside an element. The accuracy of the method is subsequently
investigated, and we compare it to well-established methods.
%
%
\subsection{Propagation of a Gaussian Wave in Vacuum} \label{sec:vacuum-wave-propagation}
\begin{figure}[!!ht]
\centering
\includegraphics[width=0.5\textwidth]{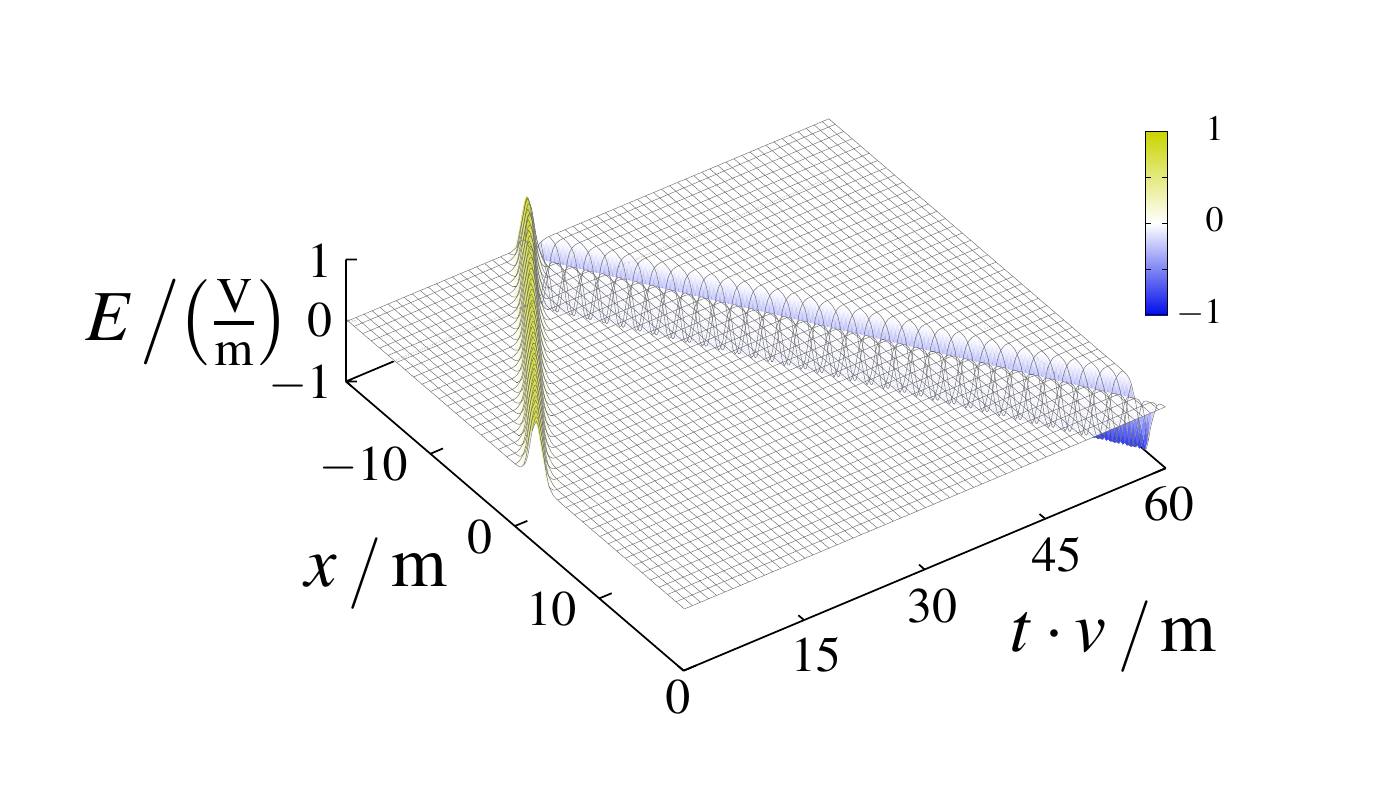}
 \caption[fig:vacuum]{Electric field of a one-dimensional Gaussian wave in a vacuum, simulated with DGT-FEM.
   The solution in the whole space-time domain of interest $(x,t) \in [-20,20] \times [0,60]$ is displayed. } \label{fig:vacuum}
\end{figure}

%
Gaussian wave propagation provides a simple test scenario for DGT-FEM. As the waveform
has high regularity, strictly increasing accuracy is expected for high approximation orders.
In \figref{fig:vacuum} we show the simulation of the propagation of a Gaussian shaped wave (initially heading leftwards). 
In this test case, we assume that the wave travels in a vacuum (i.e.~$\epsilon=\mu=1$ and $v=1$).  The cell size is normalized to unity (i.e. $\Delta_x=\Delta_t=1$) for convenience.
At the spatial boundaries of the global space-time domain, perfect electric conductor (PEC) boundary conditions are applied. \figref{fig:vacuum} shows 
the electric field of the Gaussian wave heading leftwards at an angle of 45 degrees in the space-time plane corresponding to a velocity $v=1$. 
At the boundary the wave is reflected under a sign change of the electric field while maintaining its space-time angle of 45 degrees, thus recovering the expected physical behavior.
%
\subsection{Propagation of a Gaussian Wave in Cell-Wise Homogeneous Media} \label{sec:medium-wave-propagation}
\begin{figure}[!ht]
\centering
\includegraphics[width=0.5\textwidth]{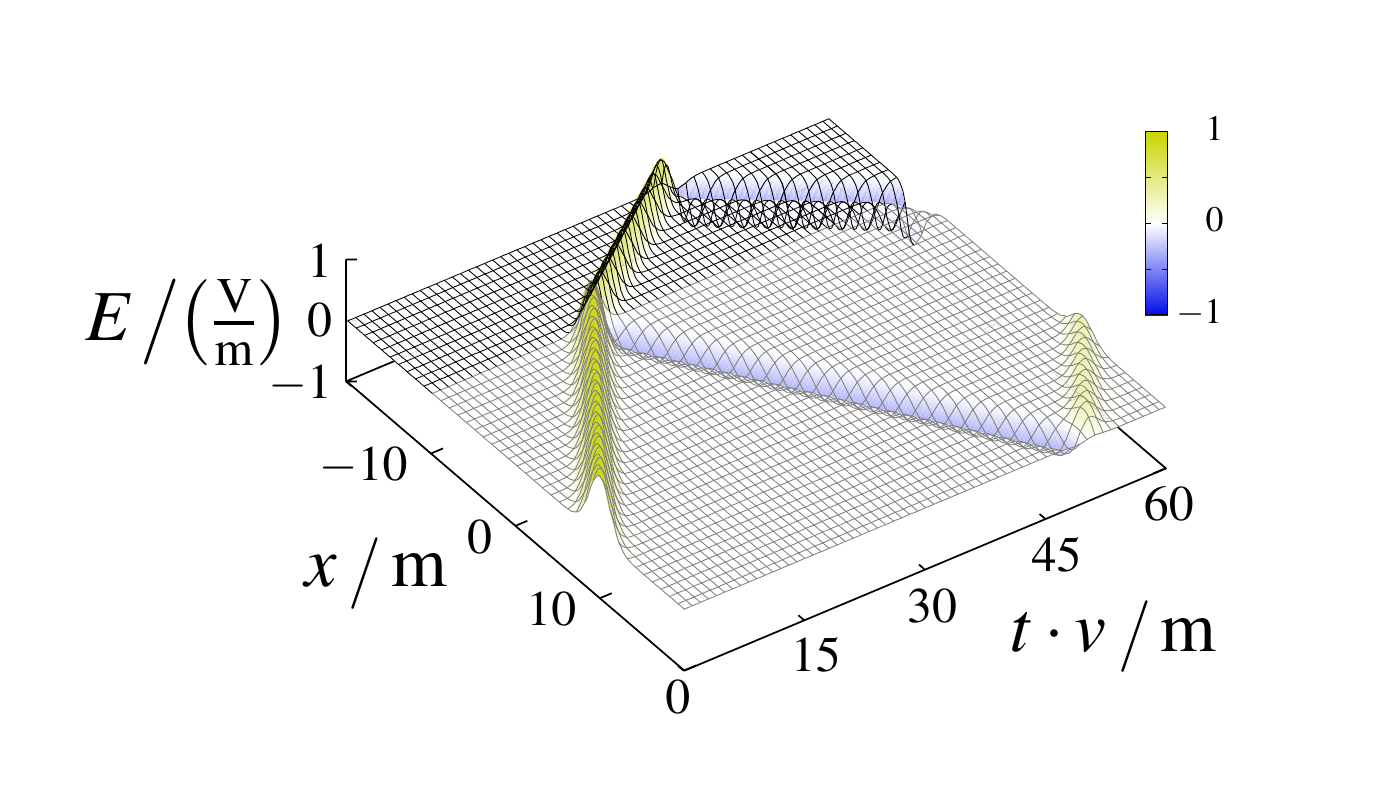}
 \caption[fig:medium]{Electric field of a one-dimensional Gaussian wave with a medium interface 
   at $x_0=-10$, simulated with the DGT-FEM. The solution in the whole space-time domain of 
   interest $(x,t) \in [-20,20] \times [0,60]$ is displayed.}  \label{fig:medium}
\end{figure}

%
We now modify the setup of the preceding section by filling the left part of the domain with
a dielectric material with $\epsilon = 4$. The medium interface is placed between two cells at $x_0=-10$.
\figref{fig:medium} illustrates the result in the space-time domain of interest.
The wave is initialized in a vacuum and heads leftwards with a space-time angle of $45$ degrees towards the dielectric material.
At time $t=20$ the wave reaches the interface and is split into a reflected and a transmitted wave. The reflected 
wave heads back rightwards preserving its $45$ degrees angle in order to be reflected at the right domain boundary.
The numerical value of the amplitude of this wave is very close to the analytical value $R=1/3$. For the transmitted wave 
the space-time angle changes to $22.5$ degrees, which corresponds to a velocity of $v_{\mathrm{med}} = 1/2$; 
the amplitude of this wave is very close to the analytical value $T=2/3$ again. Therefore, the expected reflection and 
transmission coefficients as well as wave velocities are recovered. For a wave impinging from medium to a vacuum we obtain 
$R^\prime=-1/3$ and $T^\prime=4/3$, which agrees with the expectations as well.
%
\subsection{Simulation of Inhomogeneously Filled Cells} \label{sec:innercell-wave-propagation}
\begin{figure}[!hb]
\centering
\includegraphics[width=0.5\textwidth]{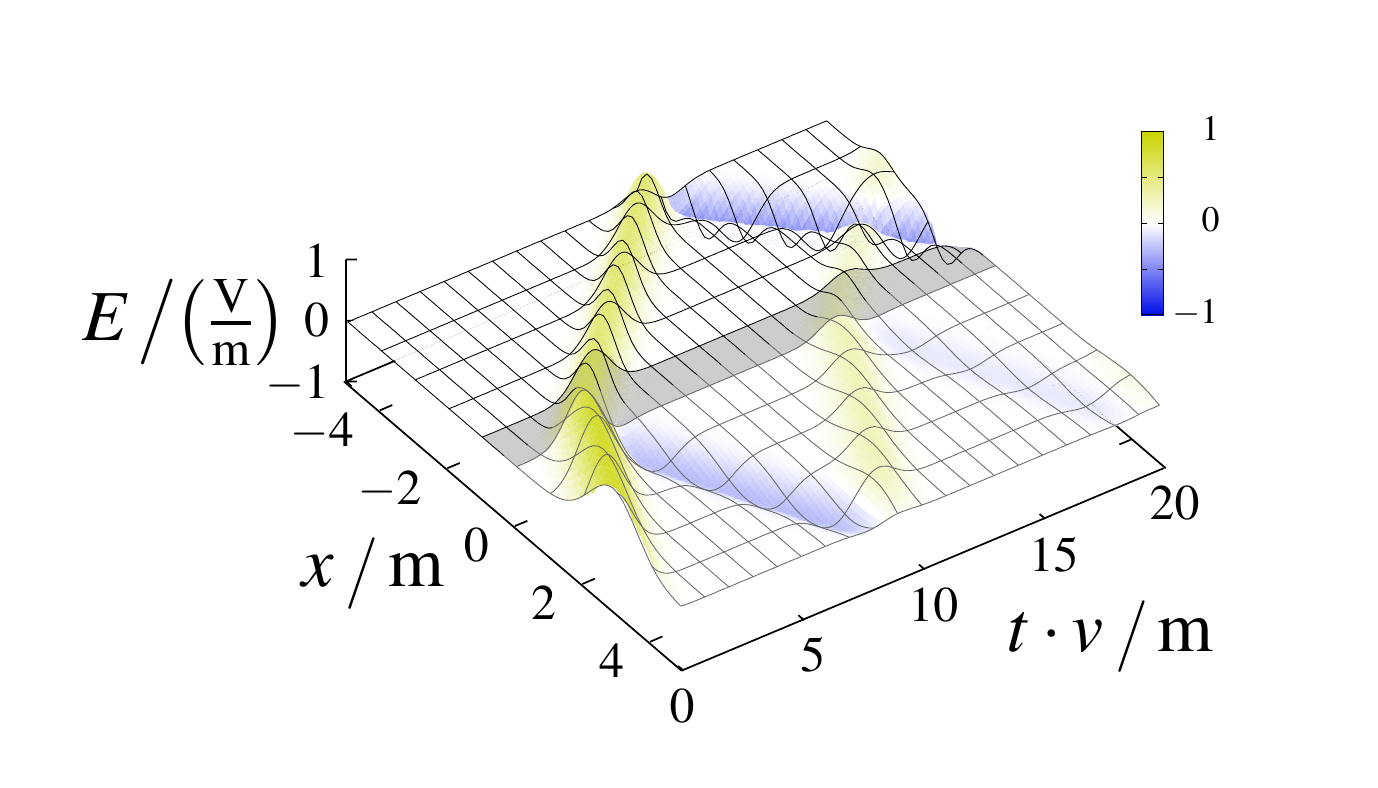}
 \caption[fig:innercell-medium]{Electric field of a one-dimensional Gaussian wave with a 
   medium interface at $x_0=-0.25$, simulated with the DGT-FEM. The material interface
  crosses an element. The solution
   in the whole space-time domain of interest $(x,t) \in [-5,5] \times [0,20]$ is displayed.} 
 \label{fig:innercell_medium}
\end{figure}

%
In \sectref{sec:Trefftz-material-interfaces} Trefftz basis functions for the case of 
material interfaces placed within a cell were derived. 
\figref{fig:innercell_medium} shows the simulation of the propagation of a wave similar to the
preceding example. While the media have exactly the same material properties as before, the medium interface 
is now placed at $x_0 = -0.25$; that is inside a cell. In this case, we also recover correct physical properties 
(i.e. reflection and transmission coefficients as well as local speed of light) as discussed before.

We note that while the preceding examples can be handled in a very similar manner
with non-Trefftz methods, this is not the case here. Typically material interfaces inside elements
are either completely avoided, or the material properties are averaged in the respective element. 
More complex methods such as immersed boundaries or cut cells are in principle capable
of handling this situation but they usually come at high numerical costs and exhibit sub-optimal
performance. Using a Trefftz approach
with the above derived basis, however, the wave behavior at the interior interface is modeled accurately. 
%
\subsection{Approximation Error and Efficiency} \label{sec:error}
%
In \sectref{sec:intro} we mentioned the advantage of Trefftz methods regarding 
accuracy compared to those using generic polynomials.
While the results presented so far are of a more qualitative nature, we perform a quantitative
error analysis in this section. To this end we first investigate the numerical error occurring during
the propagation of a Gaussian wave over a large distance in a vacuum.
%

\begin{figure}[!ht]
\centering
\includegraphics[width=0.6\textwidth]{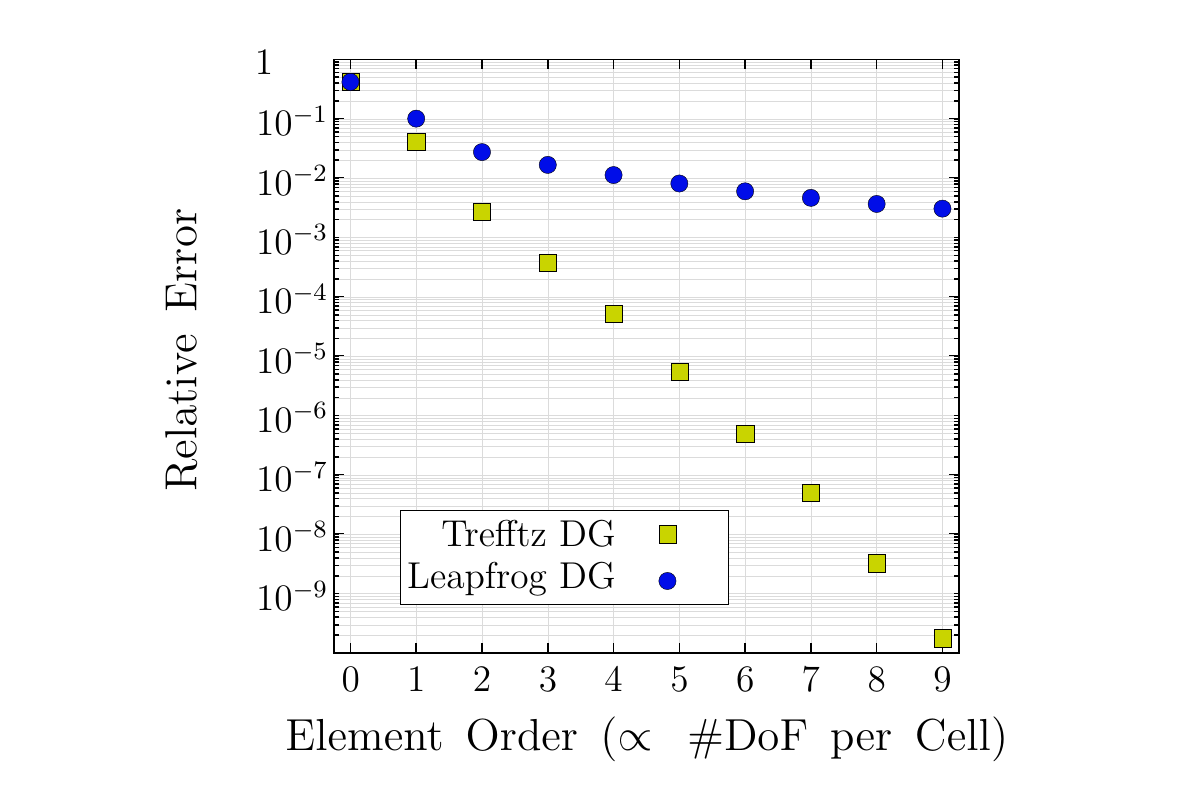}
 \caption[fig:dgvsgdt-error]{$\Lebesgue_2$ norm of the relative error for the propagation of a Gaussian wave with DGT and DGL for different polynomial orders.} \label{fig:dgvsgdt-error}
\end{figure}

\begin{figure}[!ht]
\centering
\includegraphics[width=0.6\textwidth]{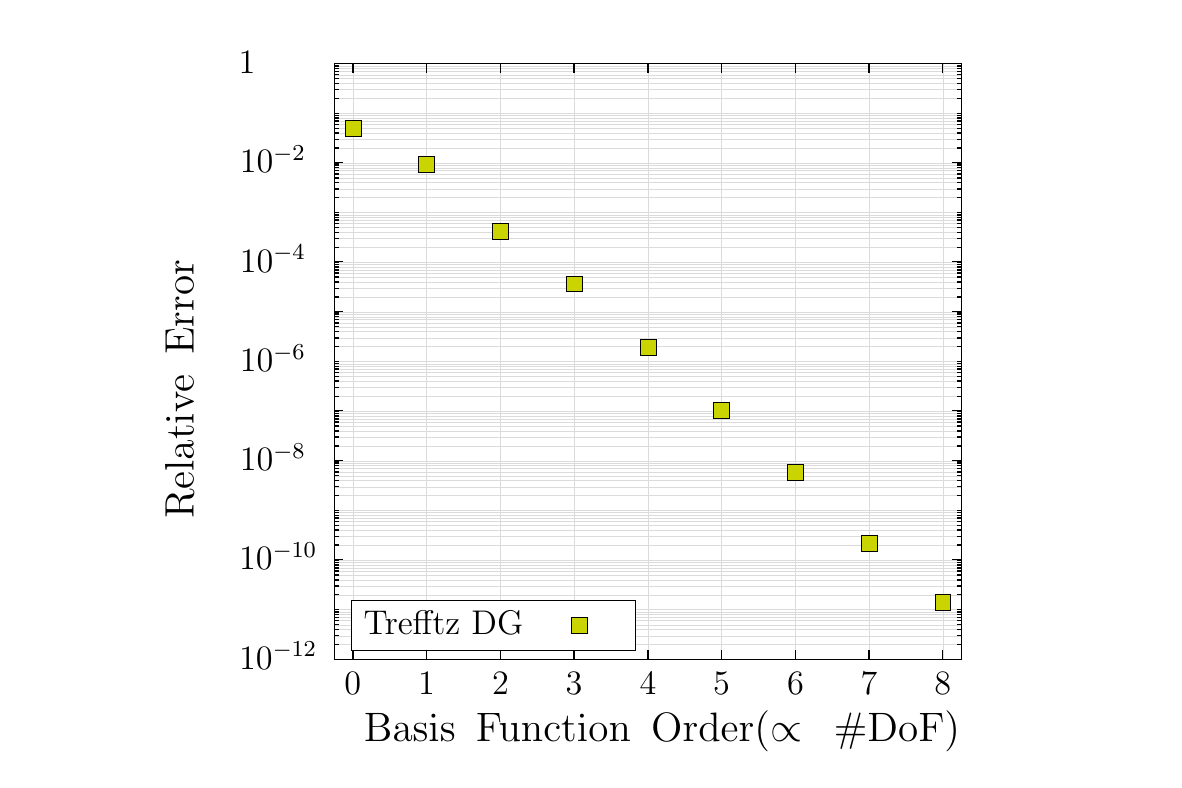}
 \caption[fig:projection-error]{$\Lebesgue_2$ norm of the relative projection error for basis functions of one DGT cell filled with two different materials as a function of the polynomial order.} \label{fig:projection-error}
\end{figure}

In \figref{fig:dgvsgdt-error} the relative ${L}_2$-error obtained with the DGT method (green markers) is compared to the error obtained using a centered DG discretization of space combined with a leapfrog time stepping scheme (DGL) \cite{Fezoui2005,Gjonaj2006} (blue markers). As can be seen from the graph, the accuracy of the DGL scheme is limited by the 
second order time integrator. However, the new DGT-FEM achieves spectral convergence 
as is inferred from the straight line in the semi-logarithmic plot. Note
that the errors indicated in the plot include spatial as well as temporal errors.
This behavior is a remarkable result of DGT-FEM. For most methods spectral convergence is only achieved in the space domain, which imposes a strong limitation on
the total error. To yield a better convergence of the global error higher order time stepping schemes
(e.g.~high order Runge-Kutta schemes) are usually employed, which increase the overall computational
costs. In DGT-FEM this is achieved at no additional computational costs. 
%
%
In \figref{fig:projection-error} we plot the relative ${L}_2$-projection-error for one cell in DGT filled with two different materials. Spectral convergence is also achieved in this case as can be inferred from the straight line in the semi-logarithmic plot again.   Therefore, spectral convergence even holds for partially filled cells in DGT.

\begin{figure}[!ht]
\centering
\includegraphics[width=0.75\textwidth]{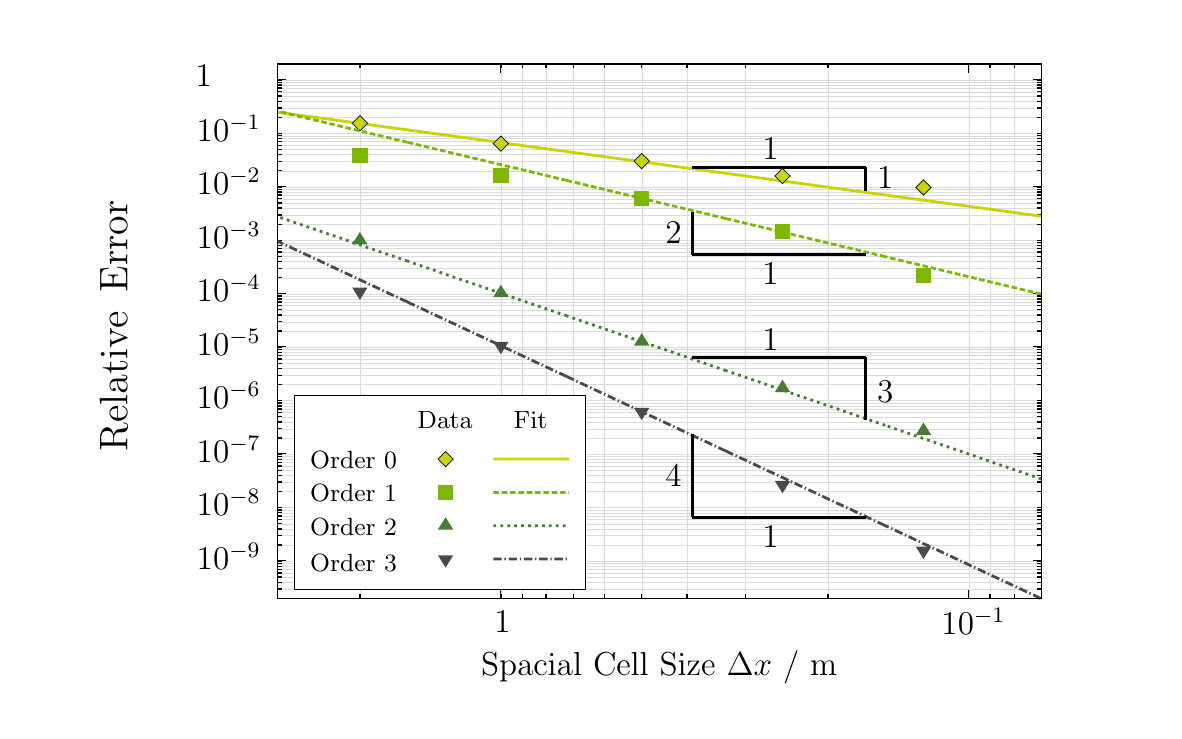}
\caption[fig:dgvsgdt-error]{$\Lebesgue_2$ norm of the relative error as obtained with DGT for different polynomial orders as a function of the spatial cell size $\Delta x$.} \label{fig:dgt_dx}
\end{figure}

\begin{figure}[!ht]
\centering
\includegraphics[width=0.75\textwidth]{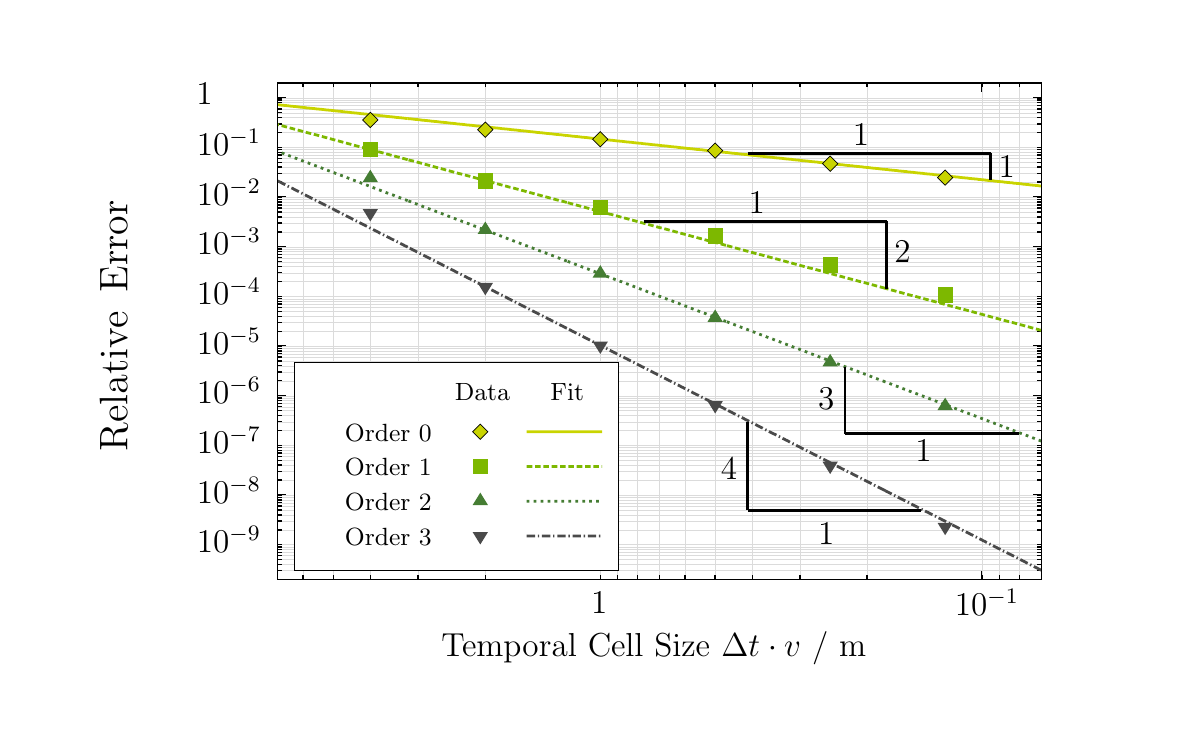}
\caption[fig:dgvsgdt-error]{$\Lebesgue_2$ norm of the relative error as obtained with DGT for different polynomial orders as a function of the temporal cell size $\Delta t$.} \label{fig:dgt_dt}
\end{figure}

Next, we will separately verify the spatial and temporal convergence orders for a fixed approximation order $p$.
To this end, we perform series of simulations where one of the grid parameters, $\Delta x$ or $\Delta t$
is successively divided in halves while the other respective parameter is fixed. This way
the convergence rates in space and time can be evaluated individually. The results are depicted in \figref{fig:dgt_dx} and \figref{fig:dgt_dt}. The plots
show that the expected convergence rates (i.e.~$p+1$ for basis functions of order $p$) are obtained for cell size variation of both $x$ and $t$.
%
%

\begin{figure}[!Ht]
 \centering
\includegraphics[width=0.75\textwidth]{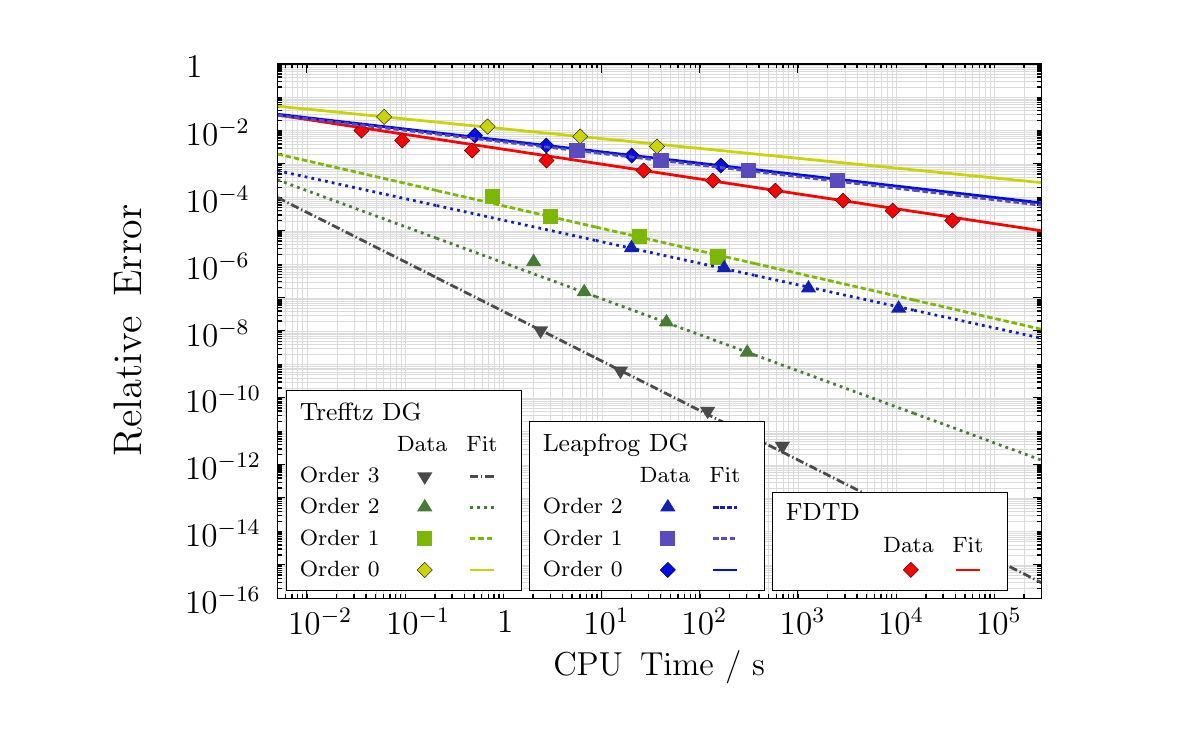}
  \caption[fig:dgvsgdt-error]{$\Lebesgue_2$ norm of the relative error for the propagation of a Gaussian wave with DGT and DGL with different polynomial orders as well as FDTD as function of CPU time.} \label{fig:efficiency}
 \end{figure}

In \figref{fig:efficiency} we plot the relative ${L}_2$ error for three methods including DGT (green curves) as a function of 
computing time. The other two methods are the DGL method mentioned above (blue curves) and a Finite Difference time domain (FDTD) method (red curve) respectively. The setup is identical for all methods, and we only show the computing time consumed by the update procedure. The error behavior of spatial DG methods (such as the DGL method used here) has been discussed in great detail in~\cite{Cohen2006,schneppphd}. One directly observes that DGT becomes the most efficient method for orders $p \ge 1$.

%
%
\section{Summary and Outlook} \label{sec:outlook}
%
We have developed a theoretical framework for the description of (electromagnetic) wave propagation problems in the time-domain
in terms of a space-time DG method. In this framework we have subsequently implemented problem-specific Trefftz-type basis functions for cells with homogeneous as well piecewise homogeneous media.
High-order time integration is an inherent property of the method, and we demonstrated spectral convergence in space-time for wave propagation problems.
As the Trefftz functions for inhomogeneous media reproduce
the exact physical behavior at media interfaces, spectral convergence is obtained even in the presence of partially filled elements. A performance 
investigation showed that the method outperforms established time-domain methods such as FDTD and DGL.

The method is applicable to higher dimensional systems (i.e. (2+1) and (3+1)),
and we are currently in the process of extending our implementation to problems governed by Maxwell's equations 
\begin{align} \label{eqn:3D-Maxwell-time}
    \mathbf{\nabla} \times  \mathbf{E} + \partial_t \big(\mu \mathbf{H} \big) \,=\, 0 \mand \nabla \times \mathbf{H} - \partial_t \big(\epsilon \mathbf{E} \big) \,=\, \mathbf{J}.
\end{align}

The fundamental difference to the (1+1)-dimensional case is that the directions of propagation are not known \textit{a priori} (i.e. there are infinitely many), as has been the case 
in the (1+1)-dimensional case (i.e. waves were only able to propagate leftwards, and rightwards). Instead, we approximate the fields locally, within each homogeneous element, by plane waves of the form
\begin{align}\label{eqn:transport-basis-3D}
  \mathbf{u}^{p} \, = \, 
  \left(\begin{array}{c}
  \mathbf{u}^{\mathbf{E},p}\\ \\\mathbf{u}^{\mathbf{H},p}
  \end{array}\right) \, = \,
  \left(\begin{array}{c}
  \mathbf{\hat{n}} \, \big( \, \mathbf{\hat{k}} \cdot \mathbf{r} \, - \, \mathbf{\hat{k}} \cdot \mathbf{v} \, t \, \big)^p \\ \\ \frac{\mathbf{\hat{k}}\times \mathbf{\hat{n}}}{Z} \, \big(\, \mathbf{\hat{k}} \cdot \mathbf{r} \, - \, \mathbf{\hat{k}} \cdot \mathbf{v} \, t \, \big)^p
  \end{array}\right),
\end{align}  
where $\mathbf{\hat{k}}$ is the direction of transport unit vector and $\mathbf{\hat{n}}$ the unit vector characterizing the polarization of the respective basis function.
Approximation of fields by plane waves is a well known and common technique. Its accuracy in the frequency domain has been studied rigorously by several research groups \cite{Huttunen:2007tq,Moiola:2011io,Cessenat:2003tl,Hiptmair2013}
but the mathematical analysis is very intricate and complex. We are not aware of similar mathematical results in the time domain and will therefore apply \eqnref{eqn:transport-basis-3D} heuristically but with a high level of confidence in its validity, supported by the numerical calculations below. It can be expected that the accuracy of plane-wave approximation for higher dimensional systems will depend on the order of the polynomial basis, on the number of directions of propagation employed and, 
to a lesser extent, on the choice of these directions (see \cite{Hiptmair2013}). More generally, approximation properties of Trefftz bases (``T-sets'') have been studied by Herrera and others (e.g. \cite{Herrera1980,Herrera2000}).

\begin{figure}[!ht]
 \centering
\includegraphics[width=0.65\textwidth]{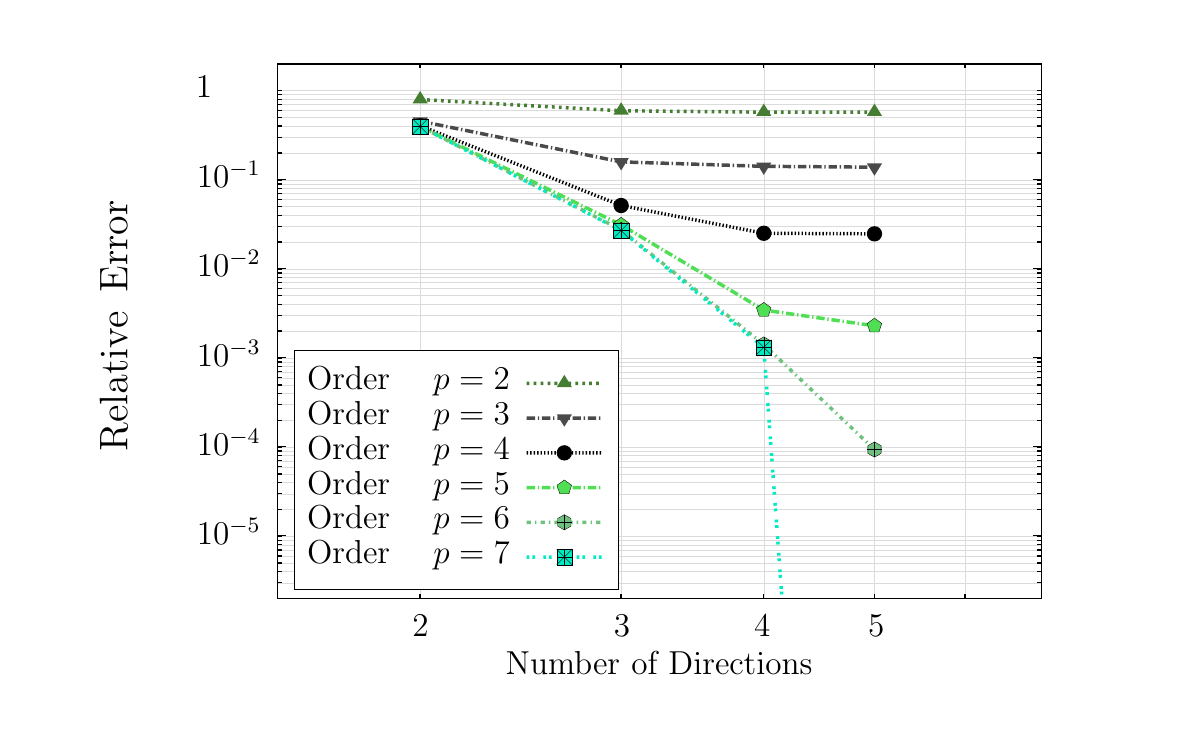}
  \caption[fig:dgvsgdt-error]{$\Lebesgue_2$ norm of the relative error of a (2+1)-dimensional seventh order initialization as function of the number of directions.} \label{fig:two-dim-order7ini}
 \end{figure}

To get some quantitative estimates, we approximated a wave of the form $\Psi (x,y,t) = \big((x-1) \cos(\alpha) + y \sin(\alpha) - t \big)^7$ 
with $\alpha = \pi /5$ using~\eqref{eqn:transport-basis-3D} with increasing order and an increasing number of directions.
The result shown in \figref{fig:two-dim-order7ini} verifies that the approximation error
depends on both parameters. In the case of five directions and order seven, the wave direction
and order is matched, and the error approaches zero. Here, we benefit from the works
on UWVF and PWDG \cite{Monk02,Huttunen:2007tq,Moiola:2011io,Hiptmair:2011p1917}
regarding the choice of directions.
%
\section*{Acknowledgments} 
%
The work of Fritz Kretzschmar is supported by the 'Excellence Initiative' of the German Federal and State Governments and the Graduate School of Computational Engineering at Technische Universit\"at Darmstadt.\\
Sascha M.~Schnepp acknowledges the support of the `Alexander von Humboldt-Foundation' through a `Feodor Lynen Research Fellowship'.\\
The work of Igor Tsukerman was supported in part by the National Science Foundation under Grant No. 1216927. Igor Tsukerman additionally thanks the Graduate School of Computational Engineering at Technische Universit\"at Darmstadt for financial support.
%
%

%
%
\end{document}